# Manufacturing of Aluminum Composite Material Using Stir Casting Process

MUHAMMAD HAYAT JOKHIO*, MUHAMMAD IBRAHIM PANHWAR**, AND MUKHTIAR ALI UNAR***

## ABSTRACT

Manufacturing of aluminum alloy based casting composite materials via stir casting is one of the prominent and economical route for development and processing of metal matrix composites materials. Properties of these materials depend upon many processing parameters and selection of matrix and reinforcements. Literature reveals that most of the researchers are using 2, 6 and 7xxx aluminum matrix reinforced with SiC particles for high strength properties whereas, insufficient information is available on reinforcement of "$Al_2O_3$" particles in 7xxx aluminum matrix. The 7xxx series aluminum matrix usually contains Cu-Zn-Mg. Therefore, the present research was conducted to investigate the effect of elemental metal such as Cu-Zn-Mg in aluminum matrix on mechanical properties of stir casting of aluminum composite materials reinforced with alpha "$Al_2O_3$" particles using simple foundry melting alloying and casting route.

The age hardening treatments were also applied to study the aging response of the aluminum matrix on strength, ductility and hardness. The experimental results indicate that aluminum matrix cast composite can be manufactured via conventional foundry method giving very good responses to the strength and ductility up to 10% "$Al_2O_3$" particles reinforced in aluminum matrix.

## 1. INTRODUCTION

The modern development in the field of science and technology demands the developments of advanced engineering materials for various engineering applications, especially in the field of transportation, aerospace and military engineering related areas. These area demands light weight high strength having good tribological properties. Such demands can only be met by development and processing of aluminum metal matrix composite materials.

The main challenge in the development and processing of engineering materials is to control the microstructure,

* Associate Professor, Department of Metallurgy & Materials Engineering, Mehran University of Engineering and Technology, Jamshoro.
** Professor, Department of Mechanical Engineering, Mehran University of Engineering and Technology, Jamshoro.
*** Professor, Department of Computer Systems Engineering, Mehran University of Engineering and Technology, Jamshoro.

mechanical properties and cost of the product through optimizing the chemical composition , processing method and heat treatment. This requires the sound theoretical and practical knowledge of the materials engineers.

The aluminum metal matrix composite materials is the combination of two or more constituents in which one is matrix and other is filler materials (reinforcements). Aluminum metal matrix may be laminated, fibers or particulates composites. These materials are usually processed through powder metallurgy route, liquid cast metal technology or by using special manufacturing process. The processing of discontinuous particulate metal matrix material involves two major processes (1) powder metallurgy route (2) liquid cast metal technology. The powder metallurgy process has its own limitation such as processing cost and size of the components. Therefore only the casting method is to be considered as the most optimum and economical route for processing of aluminum composite materials [1-25].

Literature reveals that most of the previous work was done to reinforce SiCp in various aluminum matrix composites [6,9,13,15-1, 21-24]. However, some information is available regarding $Al_2O_3$ particulates reinforced in various aluminum matrix [3,7-8,14,19-20, 22].

The effects of SiCp in Al-4.5% Cu-1.5% Mg alloy on mechanical properties of materials were investigated by Stefanos [19]. He prepared composite material using stir casting route and concludes positive response on fatigue and tensile strength in heat treated condition with addition of SiCp. The stir cast aluminum alloy matrix and process parameters were thoroughly investigated by Pai, et. al. [18]. They conclude from literature that stir casting process is relatively simple and less expansive as compared to other processing methods. They also emphasize that properties of cast composite materials depend upon uniformity of dispersoids, wetting of ceramic particles and containing low casting defects. Balasivanandha [16] investigated stirring speed and stirring time on distribution of ceramics particles in cast metal matrix composites using

SiCp reinforced in A348 aluminum matrix. They recommended that 600 rpm stirring speed and 10 minutes stirring time gave best results on properties of cast aluminum composites.

A 7075 aluminum alloy matrix reinforced with 15 volume percent of SiCp were prepared by using liquid metallurgy route by Rupa and Meenia [15]. The properties comparisons were made with parent metal. They found the improvements in mechanical properties and sliding wear resistance in aluminum cast composite materials. But they did not investigate the effect of "$Al_2O_3$" particles reinforced with cast aluminum composites with the same material.

A low cost AMMCs product having cost $2.20 per kg using rapid mixing process technique was developed by Darrel, et. al. [13], using SiCp in 359 aluminum matrix.

A10% and 20% SiCp reinforced in A535 and A 6061aluminum matrix using stir casting method was investigated by Zhou and Xu [10]. They had recommended two step mixing method and preheat treatment for achieving better mechanical properties.

The processing -microstructure- mechanical properties of aluminum base metal matrix composite materials synthesized using casting route were investigated by Gupta and Surappa [9]. They had used SiC reinforced in 6061 aluminum matrix and reported the improvements in the mechanical properties up to 15% SiCp.

Das [6] has reported up to date information regarding SiCp reinforced in aluminum metal matrix composites for engineering applications.

A similar effect of "$Al_2O_3$" particles reinforced with various aluminum matrix composites was observed by numbers of investigators [14,21-23]. They had investigated abrasive wear, mechanical properties and microstructure, using casting route and reported similar effect of "$Al_2O_3$" particles on properties of composites materials.

Literature reveals that some previous work has been conducted to investigate the effect of Cu-Zn-Mg in aluminum matrix alloy reinforced with "$Al_2O_3$" particulate using stir casting route.

Therefore, the present work was planned to investigate the effect of Cu-Zn-Mg in aluminum matrix reinforced with "$Al_2O_3$" particles using simple foundry conventional casting, for economical development of aluminum casting composite materials for engineering applications. The details of the research work are given in the subsequent sections.

## 2. EXPERIMENTAL WORK

For alloy development pure aluminum ingot, copper, magnesium and aluminum oxide average particles size 46μm were purchased from local market. The aluminum ingot was melted in a graphite crucible and alloyed with required quantity of Cu-Zn-Mg metals. The details of the theoretically selected alloy composition (design composition) and manufacturing processing are given Table 1 and Figs. 1-3.

**TABLE 1. SELECTED COMPOSITIONS OF MASTER ALLOYS IN GRAMS FOR MANUFACTURING OF COMPOSITE MATERIALS**

| Alloy No. | Cu (gm) | Mg (gm) | Zn (gm) | Al (gm) |
|-----------|---------|---------|---------|---------|
| 1. | 0.00 | 122 | 150 | 4728 |
| 2. | 100 | 140 | 300 | 4460 |
| 3. | 400 | 50 | 300 | 4250 |
| 4. | 250 | 50 | 0.00 | 4700 |
| 5. | 150 | 120 | 300 | 4430 |

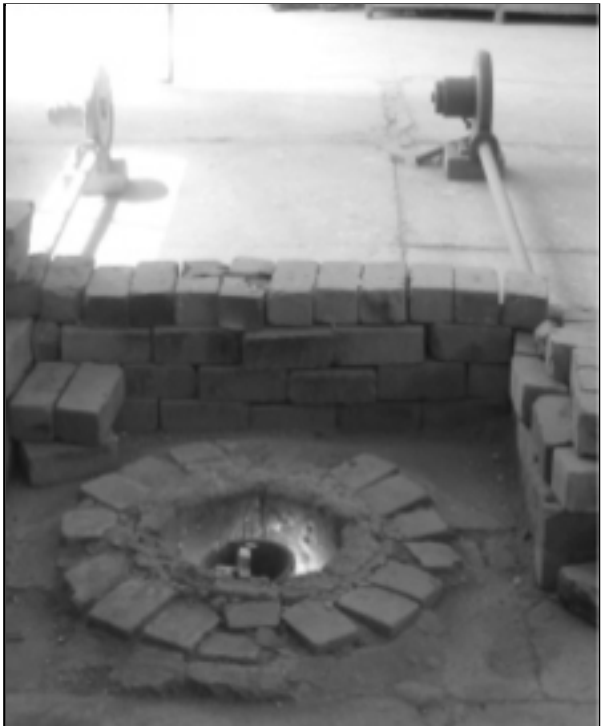

*FIG. 1. MELTING OF ALLOYS*

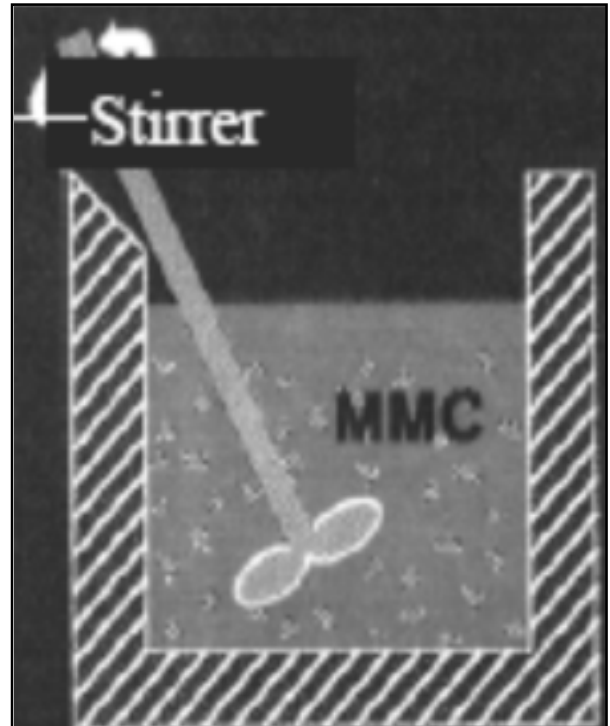

*FIG. 2. STIR CASTING PROCESS (REFERENCE FROM 25, THE SAME PROCEDURE IS USED IN PRESENT WORK)*

The developed alloys were purged with $N_2$ gas for few seconds. The required quantities of "$Al_2O_3$" particles were added in 2.5, 5, 10 and 15% weight percents. The ceramics particles were mixed in a designed mixer by stirring for few minutes and cast in a steel mold consist 25 mm dia and 130mm length (Fig. 4). The casting temperature was maintained at 750°C.

The cast samples were machined to prepare standard samples. 40 standard samples were prepared using lathe machine as shown at Figs. 5-6.

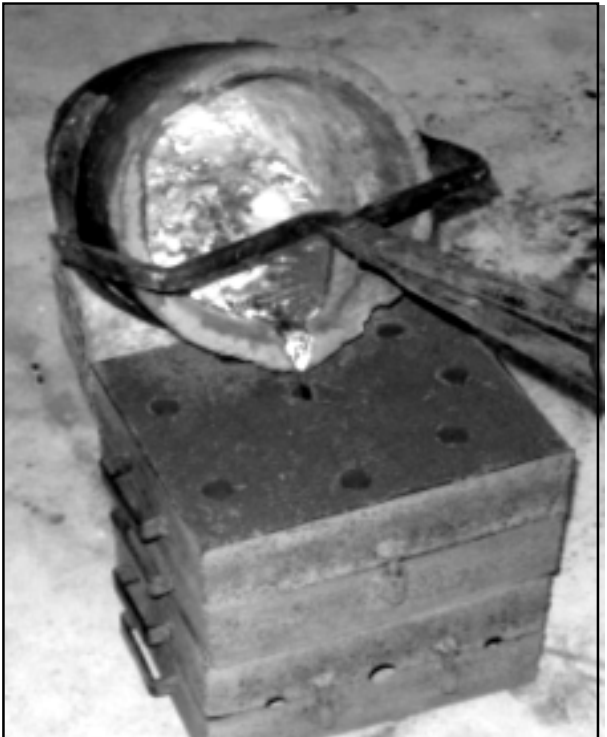

*FIG. 3. CASTING OF AMMC*

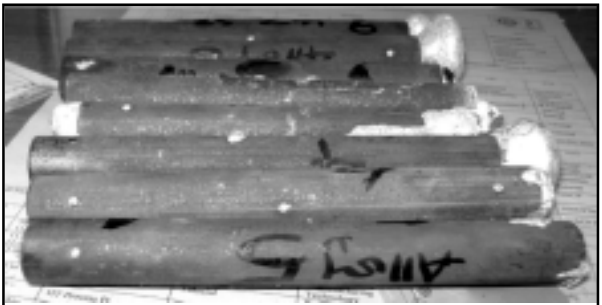

*FIG. 4. CASTING OF SAMPLES IN METAL MOLD*

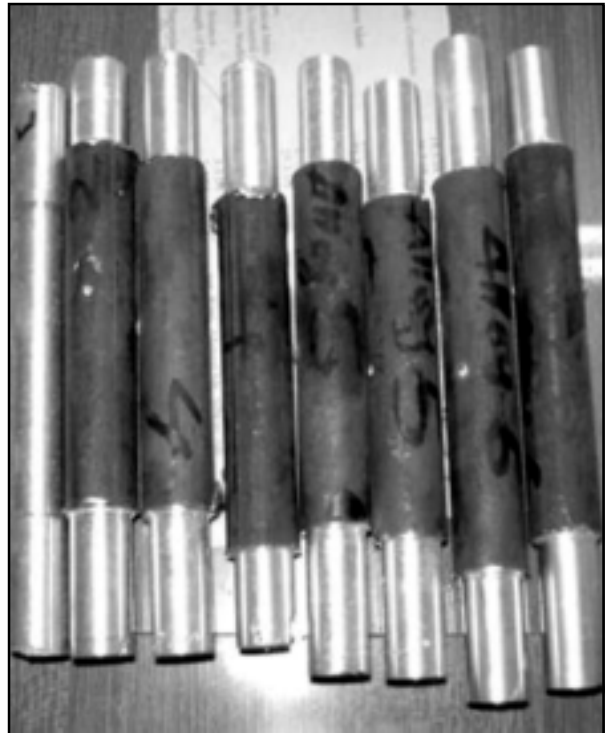

*FIG. 5. PREPARATIONS OF SAMPLES ON LATHE*

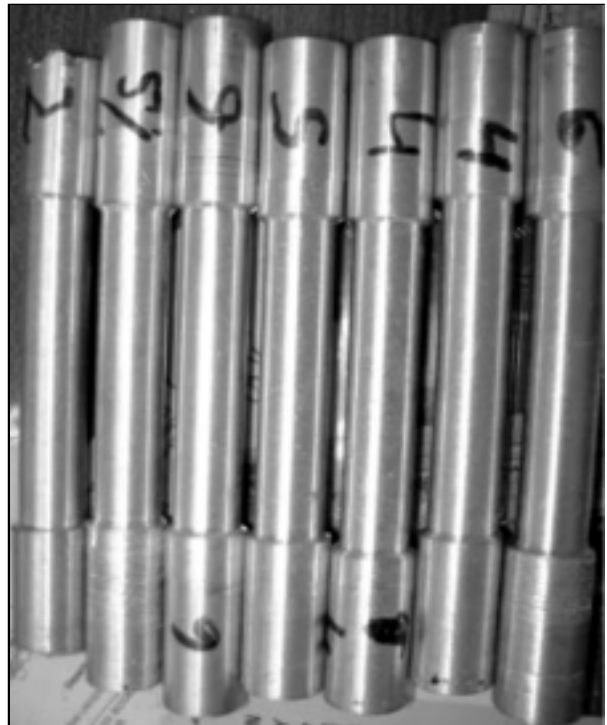

*FIG. 6. PREPARED STANDARD SPECIMENS FOR TENSILE TEST*

20 samples were solution heat treated at 580ºC for 1 hour and age hardened at 120ºC for 24 hours in a muffle furnace. Tensile strength, elongation was determined using universal tensile testing machine at Dawood College of Engineering & Technology, Karachi, Pakistan. The details of the tensile test parameters are given in Table 2.

Chemical analysis of the raw material and samples were conducted using SEM and Spectrometer analysis method as shown at Figs. 7-12 and Table 3.

The Vickers hardness tests and metallographic investigations were conducted at the Department of Metallurgy & Materials Engineering, Mehran University of Engineering & Technology, Jamshoro, Pakistan, using Vickers hardness testing machine. The Vickers hardness was converted to HRB scale. The metallographic investigations of aluminum were conducted using Scanning Electron Microscope to determine micro structural porosity Fig. 13. The Optical Metallurgical Microscope was also used to investigate aluminum matrix and ceramic particles at 100X and 200X magnification as shown in Fig. 14(a-d). Fig. 14(d) clearly showing uniformly distribution of ceramic particles.

The details of the mechanical properties such as tensile strength, hardness and elongation are given in Figs. 15-19

## 3.     RESULTS AND DISCUSSION

Keeping in view of the various limitations as reported in literature review  present research work was plan to develop aluminum alloy containing Cu-Zn-Mg in aluminum as a 7000 series  at Mehran University Jamshoro. The alloys were developed using conventional casting foundry method Figs. 1-3.  The alloys were reinforced with various %ages of "$Al_2O_3$" particles.  The alloy composition and properties were investigated using simple foundry melting alloying and conventional stir casting method.

The results mentioned in Tables 1-2 and Figs. 7-19 shows that the composition has significant effect properties of aluminum cast composites especially containing Cu-Zn-Mg in aluminum with different %age of "$Al_2O_3$" particles. The mechanical properties including tensile strength, hardness and elongation were thoroughly investigated because the literature reveals very little information on the developments of 7000 series of aluminum alloy matrix reinforced with "$Al_2O_3$" particles from elemental method using simple foundry conventional casting   method. However, most of the previous work was done in this regard is discussed as under:

Venkatraman and Suundararajan [24], conducted research on 7075 aluminum alloy and AMMCs reinforced with SiCp using powder metallurgy method. They correlate mechanically mixed layer and wear behavior with 7075 aluminum alloy and AMMCs. They observed that precipitation hardening of AMMCs gave positive response in improving in mechanical properties. However, they did not investigated the effect of "$Al_2O_3$" particles in 7075 aluminum alloy using simple conventional stir casting method.

**TABLE 2. PRAMETERS OF TENSILE TESTING MACHINE**

| Order Number | 1  2.5  Cast Sample |
|---|---|
| Charge | Rs 300 |
| Test Standard | ASTM |
| Customer | M. H JOKHIO |
| Material | Aluminium Cast Composte Material |
| Extensometer | |
| Load Cell | Maximum 250 KN |
| Pre-Load | 2  N/mm² |
| Pre-Load Speed | 10  mm/min |
| Test Speed | 50  mm/min |

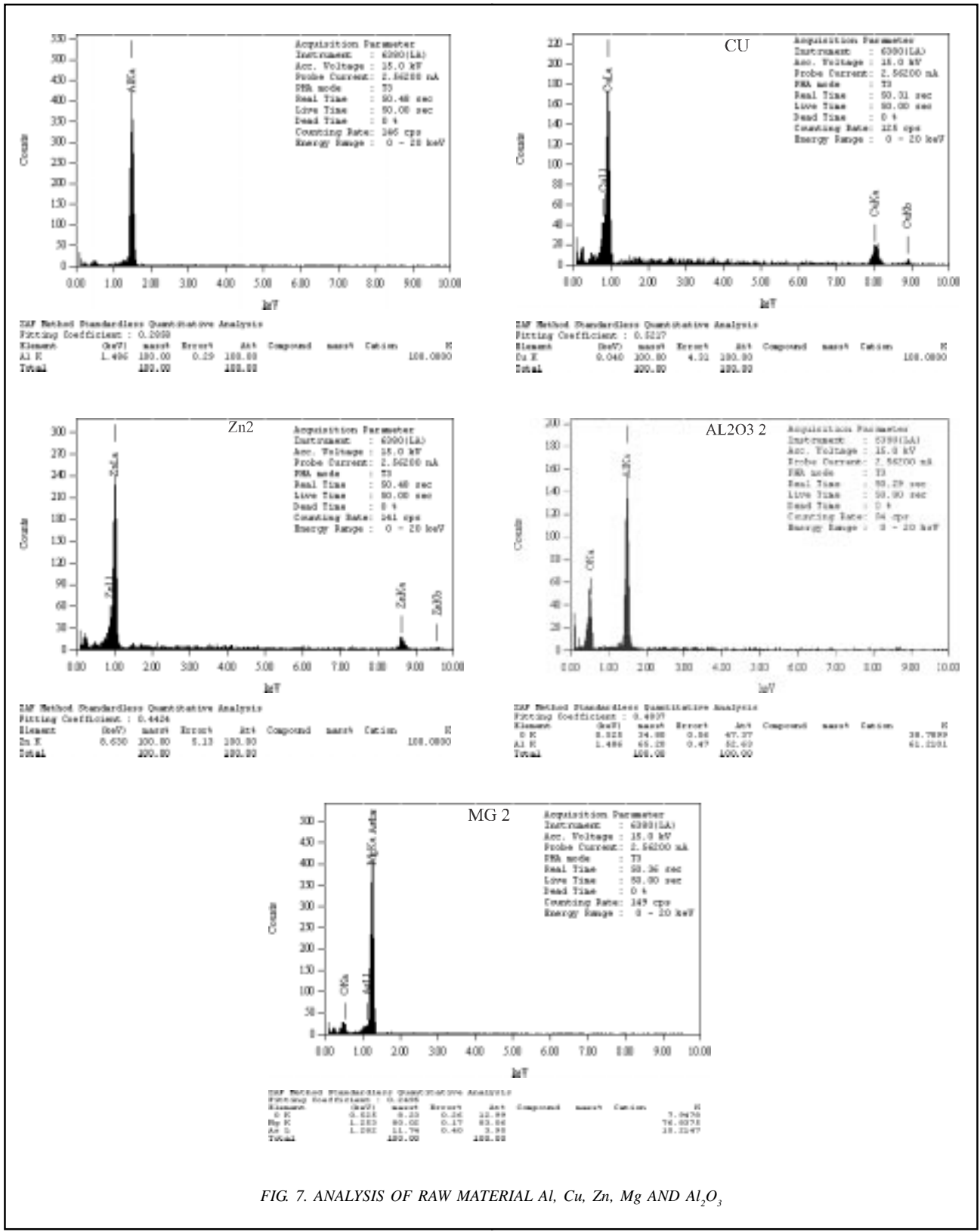

FIG. 7. ANALYSIS OF RAW MATERIAL Al, Cu, Zn, Mg AND Al₂O₃

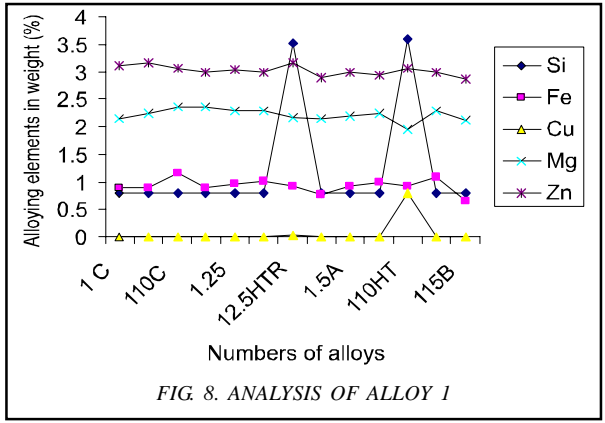

FIG. 8. ANALYSIS OF ALLOY 1

Aluminum alloy matrix containing Cu-Zn-Mg were developed by Rupa Dasgupta and Humarira Meena [15]. They developed aluminum alloy matrix from elemental metals using stir casting method and reinforced with SiCp. The properties of 7075 aluminum alloy matrix were compared with parent metal after casting and heat treatment of the same matrix the research report concluded the improvements in mechanical properties of and sliding wear resistance of cast aluminum alloy. However they did not investigated the effect of "$Al_2O_3$" particulates in 7075 aluminum matrix.

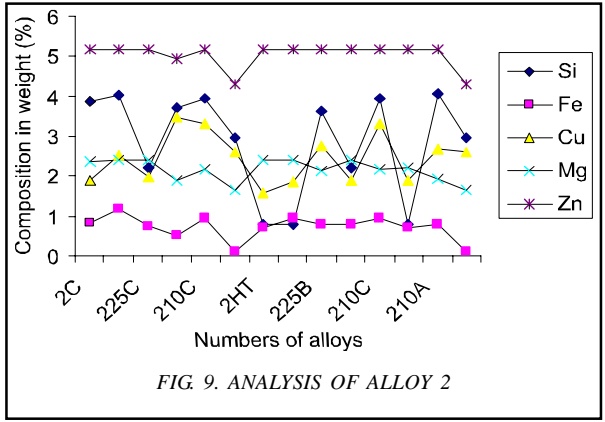

FIG. 9. ANALYSIS OF ALLOY 2

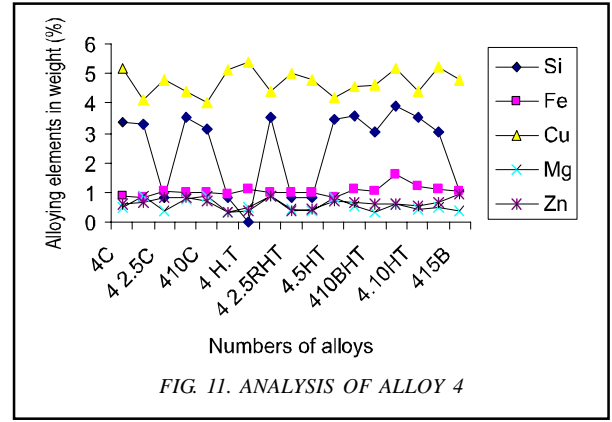

FIG. 11. ANALYSIS OF ALLOY 4

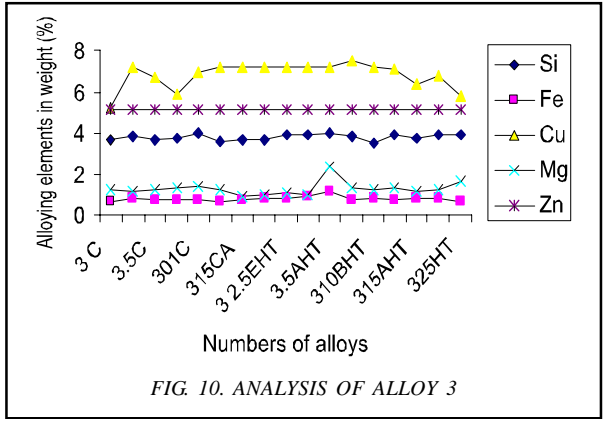

FIG. 10. ANALYSIS OF ALLOY 3

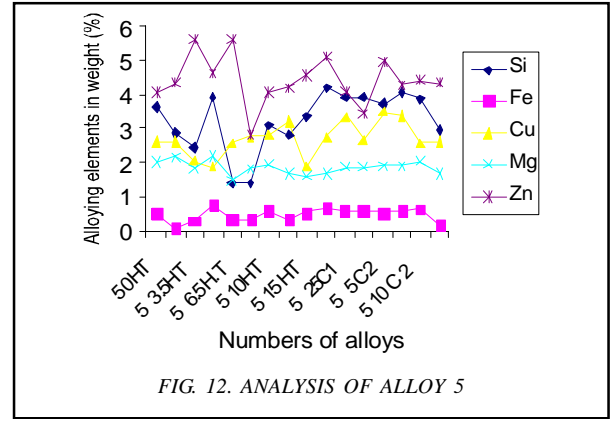

FIG. 12. ANALYSIS OF ALLOY 5

**TABLE 3. THEORETICAL COMPOSITION & ACTUAL AVERAGE COMPOSITION IN WEIGHT PERCENT AS DETERMINED BY X-RAY AND SPECTROMETER TEST RESULTS.**

| Alloy No. | Theoretical Composition of Cu (%) | Actual Composition of Cu (%) | Theoretical Composition of Mg (%) | Actual Composition of Mg (%) | Theoretical Composition of Zn (%) | Actual Composition of Zn (%) |
|---|---|---|---|---|---|---|
| 1. | 0.00 | 0 .066 | 2.50 | 2.20 | 3.10 | 3.00 |
| 2. | 2.00 | 2.45 | 2.77 | 2.15 | 5.50 | 5.00 |
| 3. | 8.00 | 7.00 | 1.00 | 1.28 | 6.00 | 5.16 |
| 4. | 5.00 | 4.70 | 0.50 | 0.60 | 0.00 | 0.60 |
| 5. | 3.00 | 2.70 | 2.00 | 1.80 | 5.00 | 4.40 |

The experimental results indicate that porosity defects were the most common factors in stir casting method Fig 18. The similar effects of porosity in stir casting were also found by numbers of investigators [1-12,14-29].

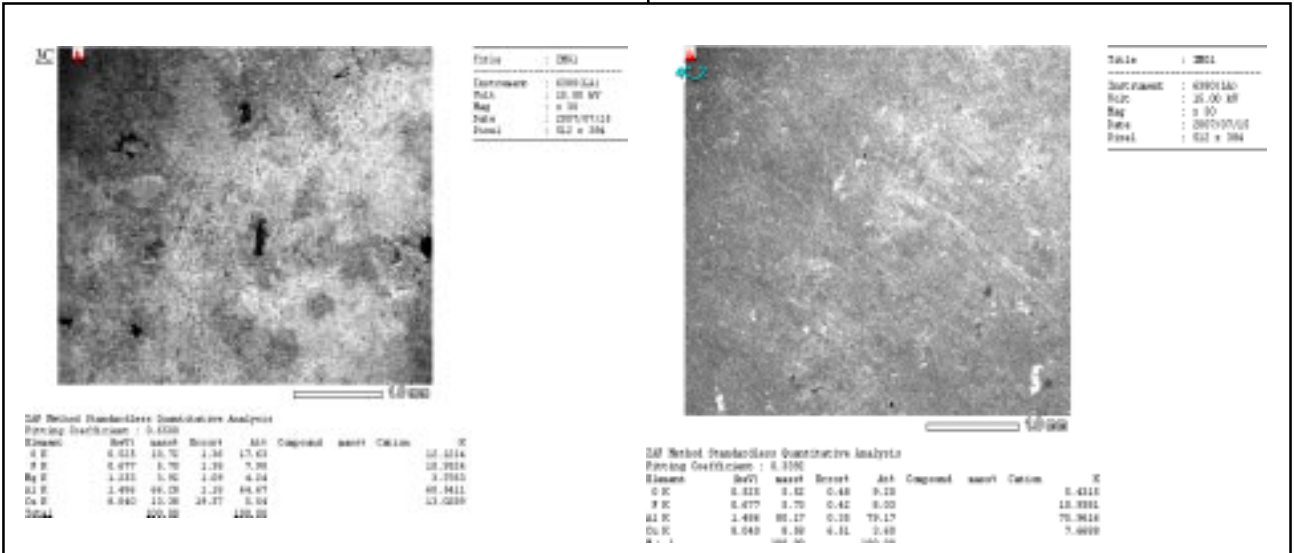

*FIG. 13 SEM MICROGRAPH FOR INVESTIGATION OF POROSITY IN STIR CAST COMPOSITES ALLOY-3 AND ALLOY-4*

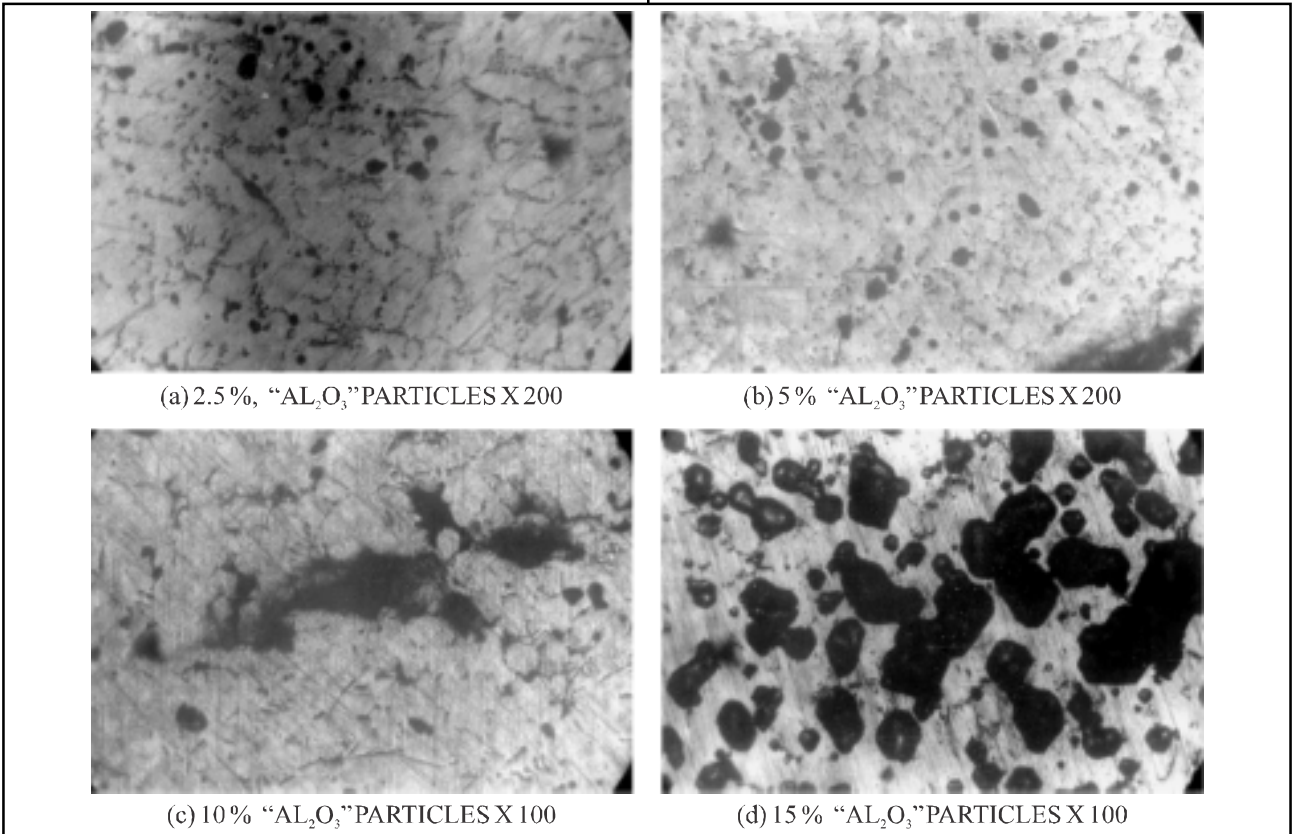

(a) 2.5 %, "AL₂O₃"PARTICLES X 200

(b) 5 % "AL₂O₃"PARTICLES X 200

(c) 10 % "AL₂O₃"PARTICLES X 100

(d) 15 % "AL₂O₃"PARTICLES X 100

*FIG. 14 A-B SHOWS THE MICROSTRUCTURE OF "Al₂O₃" PARTICLES REINFORCED IN ALUMINUM MATRIX COMPOSITES*

Comparing mechanical properties alloys contains less percentage of "$Al_2O_3$" as compared with alloy which contain the higher percentage "$Al_2O_3$" which shows some higher strength and ductility. The high percentage of "$Al_2O_3$" up to 15% in alloy 2 and 5 decreases the tensile strength. The decrease in strength of alloys is due to increase the higher percentage of "$Al_2O_3$" in alloys. Literature reveals that high "$Al_2O_3$" particles in matrix requires the higher melting, alloying treatments temperature and longer holding time is also required for homogenizing the alloy chemistry. The similar effect of the processing parameters and causes of porosity in stir cast aluminum composites were found by number of other investigators [6,7,14].

For examples Porosity decreases the mechanical properties when particle size accedes up 50µm in stir casting as reported by Taha [6].

Hashim, et. al. [5,12], reported that porosity in stir castings is produced as a result of gases entrapped in melting and during stirring/mixing, which form gas bubbles, causes large porosity. However the causes of porosity and their control are well documented in literature [6-12].

Considering the experimental results of alloy 5 which show highest tensile strength 297 MPa was due to proper wetting and uniform distribution of ceramics particles in aluminum matrix. Similar effect of distribution of ceramic particles and their proper wetting with matrix was reported by Kok [7]. They observed that large particles dispersed more uniformly than fine particles leads to agglomerations of particles and creates porosity in vertex method in production of 2024 aluminum matrix casting composite material reinforced with "$Al_2O_3$" particles. The stir casting aluminum composite material usually inherent porosity and the degree of porosity depend on the processing parameters such as pouring temperature types of matrix and volume fraction of reinforcements [6,7,14].

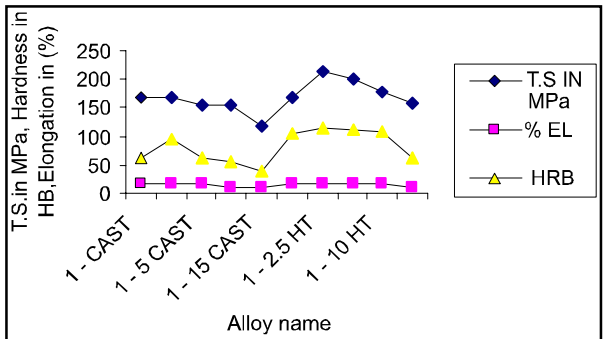

*FIG. 15. MECHANICAL PROPERTIES OF ALLY-1.*

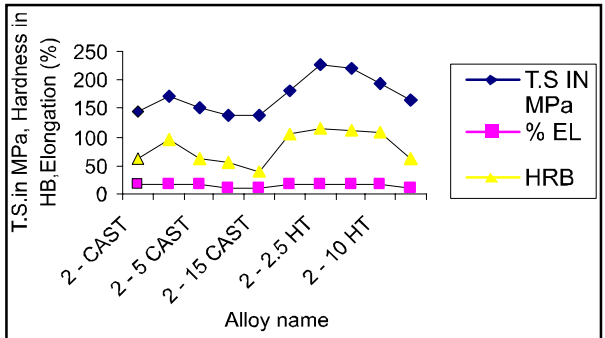

*FIG. 16. MECHANICAL PROPERTIES OF ALLOY-2*

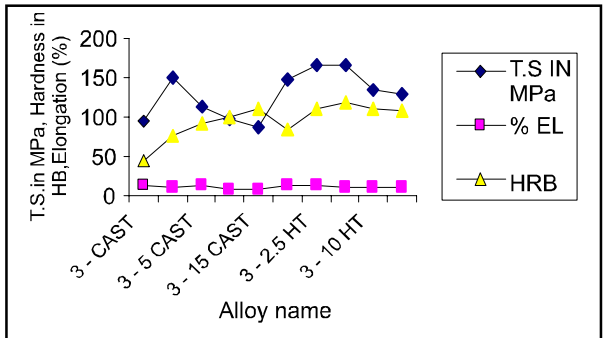

*FIG. 17. MECHANICAL PROPERTIES OF ALLY-3*

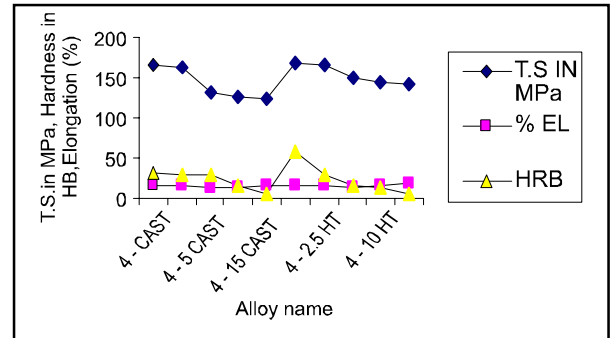

*FIG. 18. MECHANICAL PROPERTIES OF ALLY-4*

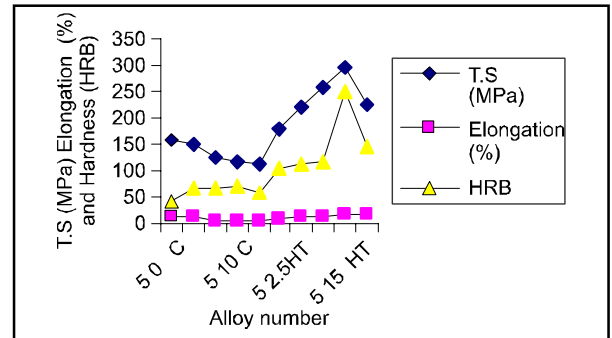

*FIG. 19. MECHANICAL PROPERTIES OF ALLOY-5*

It has been observed from the experimental work that Mg content up to 2.77% increases the wetting ability of the ceramics particles as compared to alloys which contains less or no Mg. The similar effect of Mg has been found by other investigators [3,5,11,12,15,17,22].

It is also observed that the tensile strength increases as the percentage of "$Al_2O_3$" particle increase in as cast aluminum alloys up to 2.5% "$Al_2O_3$" particles in as cast conditions. This increases is due to the presence of "$Al_2O_3$" Particles which act to refine the grain size of aluminum casting composites by nucleating fine grains size during the solidification process in as cast conditions. The decrease in tensile strength and ductility with increase the volume fraction of "$Al_2O_3$" particles up to 15% in as cast condition have been observed in aluminum alloy matrix. This is due to increase in porosity in stir casting aluminum matrix [6,7,14].

The experimental results especially in alloy 1, 2 and 5 showing good aging response on mechanical properties of aluminum alloy matrix cast composite is also yields very good response after solution treatment and ageing at 120°C for 24 hour shows increase in properties. The effect of Cu-Zn-Mg metals in aluminum matrix using SiCp were investigated by number of investigators [10-11,15,21-22]. But they did not investigate the effect of "$Al_2O_3$" particles in same aluminum matrix.

Comparing mechanical properties of present investigation with the previous work conducted by other investigators using stir casting method (Table 4), giving superior responses to strength increases up to 300 MPa and ductility up to 17 % with addition of 10% "$Al_2O_3$" particles reinforced in aluminum matrix.

## 4.  CONCLUSIONS

(i)  Aluminum alloy matrix can be developed successfully with the addition of Cu-Zn-Mg metals and reinforced with "$Al_2O_3$" using simple foundry melting alloying and casting route.

(ii)  Results show that "$Al_2O_3$" particles up to 10% increase the tensile strength 297 MPa and elongation 17% in aluminum alloy matrix containing Cu-Zn -Mg in aluminum.

**TABLE 4. COMPARISONS OF TENSILE STRENGTH AND ELONGATION RESULTS WITH THE RESULTS OF OTHER INVESTIGATORS**

| Author Name | Country | Reference | Matrix Composition | Reinforcement | Process Used | Strength | El |
|---|---|---|---|---|---|---|---|
| Jokhio | Jamshoro Pakistan | [28, 29] | Al-Cu-Zn-Mg | Particulate "$Al_2O_3$" 48μm | Stir Casting | 297MPa | 17% |
| Kok | Turkey | [7] | Al-Cu 2024 | 10% "$Al_2O_3$" Max | Vortex | 112 MPa | Not Determined |
| Yilmaza | Turkey | [14] | Si.46 MM .17 | "$Al_2O_3$" 20μm 15 Val% | Stir casting | 195 MPa | Not Determined |
| Srapppa and Rohatgi | India | [8] | Al- 11.8% Si 1-4% "$Al_2O_3$" | "$Al_2O_3$" 53-63μm | Vertex | 157Mpa with 1% "$Al_2O_3$" / 118 Mp with 4% "Al2O3" | 6% / 2.5% |
| Stefanos Skolianos | Greece | [19] | 4.5% Cu, 1.5% Mg | "$Al_2O_3$" 50μm 20 24 .2% | Stir casting | 194 MPa | Not determined |
| Azim, et all | Egypt | [21] | 2024  4.12 Cu and 1.94 Mg | "$Al_2O_3$" 50-150μm 10% "$Al_2O_3$" | Vertex | 132 MPa | 3.2% |
| Rupa Das Gupta and Humaira Meenai | India | [15] | 7075 Al-Cu-Mg-Zn Cu 1.6, Mg 2.5, Zn 5.6 | Si Cp 40 μm 15 wt% | Stir casting | As cast alloy 63 MPa T6 100 MPa | 1.13 1.56% |

(iii)   The results also suggest that Cu contents in small quantity less than 2.7% Cu increases the strength and ductility along with Mg and Zn contents in aluminum matrix as in case of alloy 1

(iv)    Porosity generally found in stir casting of aluminum composites increases with increase in "$Al_2O_3$" particles contents in aluminum matrix especially containing high %age of alloy addition .

(v)     The higher tensile strength in stir cast aluminum composites reinforced with 2.5% "$Al_2O_3$" particles in all alloy samples do not decrease the overall mechanical properties in aluminum stir casting composites  is due to refining of grain size of composite.

(vi)    Mg has pronounced effect on aluminum cast composites up to 2.77% Mg contents which increases wetability, reduces porosity and develops very good bonding with "$Al_2O_3$" particles, and aluminum matrix 1 and 5 yields superior properties as compared with other aluminum matrix 3 and 4.

(vii)   2-3% Mg in combination with 4.5% Zn in aluminum matrix reinforced with 10% of "$Al_2O_3$" alloy number 1 and 5  particles yields best combination of strength and ductility in case of a heat treated conditions.

(viii)  The over all mechanical properties can be improved by precipitation hardening process in aluminum matrix as well as in aluminum cast composites reinforced with "$Al_2O_3$" particles.

## ACKNOWLEDGMENTS

The authors greatly acknowledge the Authority of Mehran University of Engineering and Technology, Jamshoro, Pakistan, for their co-operation in providing the lab facilities and financial help for alloy developments processing and characterizations at workshop Department of Mechanical Engineering. The authors greatly acknowledge Professor Dr. Naseem, Ex Principal, Dawood College of Engineering & Technology, Karachi, Pakistan, for providing the modern material testing facilities. Thank also extended to Professor Dr. Muhammad Moazam, and Dr. Muhammad Ishaque Abro, who helped in conducting research work.

## REFERENCES

[1]     Shivatsan, T.S, Ibrahim, I.A, Mehmed, F.A., and Lavernia, E.J., "Processing Techniques for Particulate Reinforced Metal Matrix Composites", Journal of Materials Science, Volume 26, pp. 5965-5978, 1991.

[2]     Warren, H., and Hunft ,Jr., "Automotive Applications of Metal Matrix Composites", Aluminum Consultants Group, Inc Danied B. Miracle, Air Force Research Laboratory Report, pp. 1029-1032, 2004.

[3]     Ejiofor, J., and Reddy, R.G., "Developments, Processing and Properties of Particulate AL-Si Composites", Journal of Materials, pp. 31-37, November, 1997.

[4]     Taha, A.M., "Industrialization of Cast Aluminum Materials Composites", Jounral of Materials and Manufacturing Processes, pp. 618-641, September, 2001. http/ www.inform World.Com

[5]     Hashim, T., Loony, L., and Hashmi, M.S.J., "Particle Distribution in Cast Metal Matrix Composite", Jounral of Materials Processing Technology, Part-1, Volume 123, pp. 251-257, 2003.

[6]     Das, S., "Development of Aluminum Alloy Composite for Engineering Applications", Indian Institute of Materials, Volume 27, No. 4, pp. 325-334, August, 2004.

[7]     Kook, M., "Production and Mechanical Properties of "$Al_2O_3$", Particles Reinforced 2024, Al Alloy Composites", Jounral for Materials Processing Technology, Volume 161, pp. 385-387, 2003.

[8]     Surappa, M.K., Rohatgi, P.K., "Preparation and Properties of Cast Aluminum-Ceramics Particle Composites", Jounral of Materials Science, Volume 16, pp. 983-993, 1981.

[9]     Gupta, and Surappa, "Processing Microstructure Properties of Al Based Metal Matrix Composites", Journal of Key to Engineering Materials, Volume 104-107, pp. 259-274.

[10]    Zhou, W., and. Xu, Z.M., "Casting of SiCp Reinforced Metal Matrix Composites", Jounral of Materials and Processing Technology, Volume 63, pp. 358-363, 1997.

[11]    Maxim, L., Seleznew, I., James, S., A, .Cornie, All, S.Argon and Ralph, P, Mason, "Effect of Composition Particle Size and Heat Treatment on Mechanical Properties of Al-4.5 % Cu, Based, Alumina Particulates Reinforced Composites", SAE International Congress DETOIT, M I February, 23-26 1998.

[12]    Hashim, J., Loony, L., and Hashmi, M.S.J., "Metal Matrix Composites Production by Stir Casting Method", Journal of Materials Processing Technology, Volume 92-93, pp, 1-7, 1999.

[13]    Darell, R., Herling, G, Glenn J, and Wrrent, H.Jr., "Low Cost Metal Matrix composites" Journal of Advanced Materials and Process, pp. 37-40, July, 2001.

[14]    Yilmaz, and Buytoz, "Abrasive Wear of $Al_2O_3$ Reinforced Aluminum Based MMCs", Jounral of Composites Science and Technology, Volume 61, pp. 2381-2392, 2001.

[15]    Rupa, D.G., and Meenia, H., "SiC Particulate Distribution Dispersed Composites of An AL-Zn-Mg-Cu Alloy Property Comparison with Parent Alloy", Jounral of Materials Characterizations, Volume 54, pp. 438-445, 2005.

[16]    Balasivanandha, S., "Influence of Stirring Speed and Stirring Time on Distribution of Particles in Cast Metal Matrix Composites", Jounral of Material Processing Technology, Volume 171, pp. 268-273, 2006.

[17]    Pradeep, R., "Cast Aluminum Metal Matrix Composites for Automotive Applications", Journal of Materials, pp. 10-10, April, 1993.

[18]    Pai, B.C., Pllia, R.M., and Satyanaryanak, G, "Stir Cast Aluminum Alloy Matrix", Key Engineering Materials, Volume 79-80, pp. 117 128, 1993.

[19]    Stefanos S., "Mechanical Behaviour of Cast SiC Reinforced with Al 4.5% Cu-1.5% Mg Alloy", Journal of Materials Science and Engineering, Volume 210, pp. 76-82, 1996.

[20]    Mohammad, B., Ndaliman, and Akpan, P.P., "Behavior of Aluminum Alloy Castings Under Different Pouring Temperature and Speeds", Journal of Particles Technology, Volume 16, pp. 1-9, 2008 Http/ lejpt.academic drect.org/ A 11/071-080.htm1.

[21]    Azim, A.A.N., Shash, Y., Mostafa, S.F.,, and Younan, A., "Casting of 2024 Al Alloy Reinforced with "$Al_2O_3$" Particles", Journal of Materials Processing Technology, Volume 55, pp 199-205, 1995.

[22]    Redsten, A.M., Klier, E.M., Brown, A.M., Dunand, D.C., "Mechanical Properties and Microstructure of Cast Oxide-Dispersion- Strengthened Aluminum", Journal of Materials Science and Engineering, Volume A 201, pp. 88-102, 1995.

[23]    Daud, A., Abou, E.K., and Abdul, A.A.N., "Effect of "$Al_2O_3$" Particles Micro Structure and Sliding Wear of 7075 Al-Alloy Manufactured by Squeeze Casting Method", Journal of Materials Engineering and Performance, Volume 13, pp. 135-143, 2004.

[24]    Venkatararaman, and Sundararajan, "Correlation Between the Characteristics of the Mechanically Mixed Layer and Wear Behavior of Aluminum AL-7075 Alloy and AL7075 Alloy and AL -MMCs", Jounral of Wear, Volume 245, pp. 22-28, 2004.

[25]    Elmas, S., "Aging Behavior of Spray Cast AL-Zn-Mg-Cu Alloys", Turkish Jounral of Engineering & Environmental Science, Volume 25, pp. 68-1-686, 2001.

[26]    Shabestari, and Moemeni, "Effect of Copper and Solidification Conditions on the Microstructure and Mechanical Properties of Al-Si-Mg Alloys", Journal of Materials Processing Technology, Volume 153-154, pp. 193-198, 2004

[27]    Hieu and Nguyen, "The Effects of Solidification Rates on Porosity Formation and Cast Microstructure of Aluminum Alloy A356 Laboratory Module 3", EGR250 Materals Science & Engineering, February, 2005.

[28]    Jokhio, M.H., Panhawar, M.I., and Unar, M.A., "Modeling Mechanical Properties of Cast Aluminum Alloys", Mehran University Research Journal of Engineering & Technology, Volume 28, No. 3, Jamshoro, Pakistan, July, 2009.

[29]    Jokhio, M.H., "Modeling of High Strength and Wear Resistance Aluminum Alloy Based Casting Composite Material", Ph.D. Thesis, Mehran University of Engineering & Technology Jamshoro, Sindh, Pakistan, March, 2010.